\documentclass[acmlarge]{acmart}

\usepackage{gensymb}
\usepackage{siunitx}
\usepackage{wrapfig}
\usepackage{etoolbox}
\usepackage{multirow}
\usepackage{soul}
\usepackage{caption}
\BeforeBeginEnvironment{wrapfigure}{\setlength{\intextsep}{0pt}}

\newcommand{\dps}{{\si[per-mode=symbol]{\degree\per\second}}}
\newcommand{\ms}{{\si{ms}}}
\newcommand{\s}{{\si{s}}}
\newcommand{\Hz}{{\si{Hz}}}
\newcommand{\cm}{{\si{cm}}}

\newcommand{\refFig}[1]{Figure~\ref{fig:#1}}
\newcommand{\refEq}[1]{Equation~\ref{eq:#1}}
\newcommand{\refTbl}[1]{Table~\ref{tbl:#1}}
\newcommand{\refSec}[1]{Section~\ref{sec:#1}}
\newcommand{\revcorr}[1]{#1}

\def\myfigure#1#2{\begin{figure}[t]\centering\includegraphics*[width = \linewidth]{#1}\caption{#2 }\label{fig:#1}\end{figure}}

\def\myfigurelong#1#2{\begin{figure}[tp]\centering\includegraphics*[width=\textwidth,height=0.85\textheight,keepaspectratio]{#1}\vspace{0pt}\caption{#2 }\label{fig:#1}\end{figure}}
\newcommand*\mean[1]{\overline{#1}}

\AtBeginDocument{%
  \providecommand\BibTeX{{%
    \normalfont B\kern-0.5em{\scshape i\kern-0.25em b}\kern-0.8em\TeX}}}

\begin{document}

\title{Practical Saccade Prediction for Head-Mounted Displays: Towards a Comprehensive Model}

\author{Elena Arabadzhiyska}
\affiliation{%
  \institution{Max-Planck-Institut für Informatik}
  \country{Germany}
}

\author{Cara Tursun}
\affiliation{%
  \institution{Università della Svizzera italiana}
  \country{Switzerland}
}
\affiliation{%
	\institution{University of Groningen}
	\country{Netherlands}
}

\author{Hans-Peter Seidel}
\affiliation{%
  \institution{Max-Planck-Institut für Informatik}
  \country{Germany}
}

\author{Piotr Didyk}
\affiliation{%
  \institution{Università della Svizzera italiana}
  \country{Switzerland}
}

\begin{abstract}
    Eye-tracking technology has started to become an integral component of new display devices such as virtual and augmented reality headsets. Applications of gaze information range from new interaction techniques exploiting eye patterns to gaze-contingent digital content creation. However, system latency is still a significant issue in many of these applications because it breaks the synchronization between the current and measured gaze positions. Consequently, it may lead to unwanted visual artifacts and degradation of user experience. In this work, we focus on foveated rendering applications where the quality of an image is reduced towards the periphery for computational savings. In foveated rendering, the presence of system latency leads to delayed updates to the rendered frame, making the quality degradation visible to the user. To address this issue and to combat system latency, recent work proposes to use saccade landing position prediction to extrapolate the gaze information from delayed eye tracking samples. While the benefits of such a strategy have already been demonstrated, the solutions range from simple and efficient ones, which make several assumptions about the saccadic eye movements, to more complex and costly ones, which use machine learning techniques. Yet, it is unclear to what extent the prediction can benefit from accounting for additional factors and how the more complex predictions can be performed efficiently to respect the latency requirements. \revcorr{This paper presents a series of experiments investigating the importance of different factors for saccades prediction in common virtual and augmented reality applications. In particular, we investigate the effects of saccade orientation in 3D space and smooth pursuit eye-motion (SPEM) and how their influence compares to the variability across users. We also present a simple yet efficient correction method that adapts the existing saccade prediction methods to handle these factors without performing extensive data collection.}
\end{abstract}



\maketitle

\section{Introduction}

\revcorr{Novel head-mounted devices offer new and exciting ways of interacting with physical and virtual content. At the same time, these devices pose new challenges regarding human-computer interaction and content creation methods. For example, the use of standard interaction tools such as a computer mouse or a keyboard becomes less practical as displays obstruct the entire field of view of the observer. The wide-field-of-view capabilities also significantly increase the quality demands. It becomes common that new head-mounted displays require 8K+ rendering resolution for each eye at 90+\,Hz framerate\footnote{\url{https://varjo.com/products/aero/}}. Key enablers for the novel interaction and efficient rendering techniques are eye trackers that are being incorporated into the recent designs of head-mounted displays \cite{hua2006}. The precise information about the gaze location not only enables new interaction techniques \cite{majaranta2014} but also opens many opportunities for gaze-contingent display techniques that optimize image generation to provide higher quality at a reduced computational cost. A particular example of such a technique is foveated rendering \cite{guenter2012, patney2016, swafford2016, meng2018, tursun2019}, which reduces the use of computational resources by degrading the image quality for peripheral vision, where the human visual system is less sensitive to image distortions.}

Gaze-contingent techniques highly rely on accurate and instantly updated gaze-location prediction. An inaccurate prediction may lead to unintended content presented to the users or suboptimal visual quality, and as a consequence, a degraded user experience. There are several factors influencing the accuracy of the eye-tracking information. In this work, we focus on the latency, which is a result of both hardware and software limitations in a gaze-contingent display \cite{Stein2021}. \revcorr{In the context of foveated rendering, whose goal is to produce high-quality content only for the fovea, the system latency leads to delayed updates of the high-quality image region. This problem is critical for a short period following fast eye movements called saccades. Shortly after the saccade ends, the available gaze prediction is delayed due to the system's latency, and the fovea is exposed to low-quality content. Once the prediction quality stabilizes, the foveated rendering updates the image quality in the fovea region. Both the lower quality content after the saccade and the late quality change can be often observed by a viewer leading to characteristic popping artifacts.} Recently, Arabadzhiyska et al.~\shortcite{Arabadzhiyska2017} proposed a technique to limit this undesired effect. To perform the quality update of foveated rendering ahead of time, they leverage the saccadic suppression effect, which is the reduced sensitivity of the human visual system during the saccade. To this end, they develop a prediction method that is based on few initial eye-tracker samples to predict the saccade landing position. With the help of such a technique, when the saccade ends, the high-quality foveal rendering is positioned correctly and no popping artifacts are observed.

The success of such a technique is mainly dependent on the accuracy and efficiency of saccade landing position prediction. The method by Arabadzhiyska et al.~\shortcite{Arabadzhiyska2017} is computationally efficient, but it relies on the assumption that the saccade displacement profile depends solely on the saccade's length. Other techniques, such as the machine-learning-based approaches by Morales et al.~\shortcite{morales2018, morales2021}, define saccades in 2D screen space, but do not account for the fact that saccades are often combined with vergence eye movements. Additionally, the network inference required for the prediction is too expensive to use them in real-time foveated rendering applications. 

\revcorr{This work goes beyond existing models for predicting the saccade landing position by investigating additional factors that affect these eye movements and their prediction. More specifically, we focus on dynamic scenarios in VR and AR devices where saccades are combined with vergence movements and smooth pursuit eye motion (SPEM). We design and conduct user experiments that measure saccade profiles in such scenarios. Several previous works, for example \cite{Collewijn1988_vert,Collewijn1988_hor}, have already conducted similar experiments using accurate eye trackers, such as these using the scleral search coil technique. These studies have demonstrated the impact of the additional factors on the saccade profiles. Compared to them, we do not provide new insights into the physiological characteristic of saccades. Instead, we analyze these factors in the context of practical applications of saccade prediction techniques in VR and AR scenarios. For this reason, we also refrain from using coil-based eye trackers and opt for optical solutions, which despite their lower accuracy, are the most suitable solution. To our knowledge, this work is the first to investigate the impact of such factors as vergence or SPEM on the saccade prediction in current VR and AR devices. Additionally, we propose a method for correcting existing models for prediction to account for these factors. To summarize, the main contributions of this paper are the evaluation of the influence of saccade orientation in 3D space and SPEM on the saccade prediction techniques and a simple yet efficient method for adapting the existing prediction technique \cite{Arabadzhiyska2017} to handle these factors and for customizing the prediction. We believe that our investigation and technique will also help future developments of machine-learning-based techniques by limiting the required amount of training data. } 

\section{Background and Related work}
\label{sec:related_work}

    \subsection{Saccade characteristics}
    \label{sec:saccade_characteristics}
    	Saccades are simultaneous movements of both eyes to shift the gaze direction towards the visual stimulus that is away from the point of fixation \cite{schor2011, leigh2015}. They are characterized by a rapid acceleration until the maximum velocity is reached, and then a deceleration to full stop, typically followed by corrective small eye movements around the target \cite{westheimer1954}.
    	
    	\paragraph{Pre-programmed behavior} 
    	Although it is possible to observe saccades up to $100\degree$ amplitude, humans perform small saccades more frequently than large ones under natural viewing conditions \cite{bahill1975b}. Consequently, most of saccades last a brief amount of time ($<70\ms$), that is approximately equal to the time it takes for visual information to reach the brain's ocular motor mechanisms \cite{leigh2015}. Therefore, saccades exhibit a pre-programmed behavior and visual stimuli has negligible effect on a saccade when presented in the last $80$-$100\ms$ preceding the saccade onset or during a saccade \cite{young1963, becker1979}.
    	
    	\paragraph{Factors that affect the velocity} 
    	The velocity of a saccade is affected by multiple factors such as the source and target positions in the visual field, as well as the orientation of their trajectory (e.g., nasal vs temporal) and most notably by the distance between the source and target (i.e., the amplitude of the saccade) \cite{boghen1974}. Initial attempts for studying saccades revealed a relationship of the saccade amplitude to its duration and the peak velocity (a.k.a. the main sequence). It is commonly observed that the duration of the saccades show a nonlinear increase up to approximately $5\degree$, where it starts increasing linearly with the saccadic amplitude \cite{bahill1975, carpenter1988}. Similarly, the peak velocity increases linearly with saccadic amplitude up to $15\degree$--$20\degree$, where it reaches a saturation limit at approximately $600\dps$--$800\dps$ \cite{bahill1975}. The initial eye position and the orientation of trajectory also affect their velocity. The saccades which start at the periphery of the orbit and directed towards primary orbital position (centripetal) are on average faster than the saccades performed in the opposite direction (centrifugal) \cite{pelisson1988}. Similarly, the saccades performed in the horizontal direction reach higher peak velocities than those performed in the vertical direction; however, the difference becomes less significant for older adults \cite{irving2019}. There are some studies which show that the viewed content has an effect on the velocity profiles such that saccades may deviate from a velocity profile that can otherwise be modeled with a compressed exponential model \cite{costela2019, han2013}.
    	
    	\paragraph{Interactions with vergence}
    	Vergence is the slow ($ \approx \mkern-5mu 0.5 - 1\s$) movement of eyes in the opposite directions when the change in the binocular fixation involves a change in depth. If the change in the visual direction also accompanies a change in the depth, saccade and vergence take place simultaneously. In case of such combined saccade and vergence movements, they interact with each other \cite{ono1978}. Although saccade takes significantly shorter time ($ \approx \mkern-5mu 50\ms $) than vergence, a large portion ($ 40-100\%$) of vergence takes place during the saccade when they are combined \cite{enright1984, enright1986}. This shows an effective ``mediation'' of vergence by saccades. A closer inspection of peak velocities reveals that vergence speeds up while saccade slows down when they are combined \cite{erkelens1989, collewijn1997, yang2004}. However, the combined eye movement is completed in a shorter duration of time. The speeding up of vergence is observed during both horizontal and vertical saccades \cite{zee1992}. However, although combined eye movements are faster than pure vergence, the latency until the onset of eye movements is increased by $18$--$30\ms$ \cite{yang2002}. In addition, the accuracy of saccades is reduced and corrective saccades are required more often when they are combined with vergence \cite{yang2004}.
        		
    	
    	\paragraph{Saccades towards stationary targets during smooth pursuit eye movements}
    	The saccades and SPEMs are known to interact with each other. However, experimental data shows that saccades do not add up linearly with SPEMs \cite{jurgens1975}. On the contrary, for the saccades performed during an ongoing SPEM, the velocity of smooth pursuit is reduced before and after saccades performed in the opposite direction and after saccades performed in the same direction as the pursuit. The decrease is depends on the saccadic amplitude.
    	 
    	In order to describe the neural saccade programming process during SPEMs, two types of positional error vectors are defined that may explain the planned amplitude and direction of the saccade; namely, based on \emph{retinal error} and based on \emph{spatial error} \cite{mckenzie1986}. When a target is briefly flashed during a SPEM, the eyes stay in the smooth pursuit until the onset of the saccade for approximately $100$--$200\ms$. During this brief amount of time the position of the eyes change with respect the initial position when the target was flashed. If the saccades are planned based on the retinal error between source and target positions, then the neural programming of saccades would take place according to the displacement vector between the initial position and target position without taking into account the displacement of eyes until the onset of the saccade. On the other hand, the saccades programmed according to spatial error would compensate for the displacement of eyes between the initial position and the onset of the saccade. The type of positional error used by the brain to plan saccades determines the accuracy of the saccade. Initial experiments on this matter were contradictive and inconclusive. Some studies showed a correlation of saccades with retinal error \cite{mckenzie1986}. Others studies showed a correlation with spatial error if the flash is presented for a longer amount of time \cite{schlag1990, herter1998}. An explanation for the differences between the results obtained from these experiments is that the perceived motion of the target might be playing a role in the saccadic accuracy \cite{zivotofsky1996}. When the target velocity is taken into account saccades are correlated with retinal error measured at the moment of target step \cite{smeets2000}. The studies on humans show a great variability in the accuracy of saccades during SPEM  \cite{gellman1992, baker2003}. However, the source of poor localization is not well understood because there was no correlation found between the pursuit velocity ($15\dps$, $30\dps$, and $45\dps$) and the amount of saccadic inaccuracy \cite{ohtsuka1994}.

    \subsection{Saccadic suppression}
    \label{sec:saccadic_supression}
    	The image of the real world rapidly shifts across the retina during a saccade. Yet, we do not observe motion blur in the image we perceive due to reduced visual sensitivity \cite{ditchburn1955}. The duration of the reduced sensitivity spans a time interval that starts as soon as $40\ms$ before the start of the saccade and lasts up to $80\ms$ after it ends \cite{volkmann1962, latour1962, bouman1965, zuber1966}. The suppression is characterized by a selective suppression of lower spatial frequencies and the suppression effect decreases as the spatial frequency of the stimulus increases \cite{volkmann1978, burr1994}. In addition to the reduction in spatial contrast sensitivity, the target position information is also suppressed \cite{beeler1967}. However, the saccadic suppression does not result in perceiving a visual ``black-out'' due to the visual persistence of retinal images before the saccade \cite{ritter1976, campbell1978}. \revcorr{Electrophysiological studies on primates identified a reduction of neural responses just before and during saccades, followed by amplified responses and enhancement of neural signaling in the post-saccadic phase \cite{IBBOTSON2009R493}.}
    	
    	Despite the reduced visual sensitivity during the saccades, intra-saccadic perception is still possible. When the saccade peak velocity approximately matches the velocity of sinusoidal gratings rapidly drifting in the same direction, it results in perceived static image of the stimulus for a very brief amount of time during the saccade \cite{deubel1987}. Stimulus motion, which is otherwise imperceptible during fixations, can be also perceived during saccades especially when the combined movements of the stimulus and eyes result in retinal frequencies between $10$--$25\Hz$ \cite{castet2000}. Based on intra-saccadic perception, an important question is whether saccadic suppression is just a consequence of motion blur in the retinal image or not. Recent studies show that saccadic suppression is not just as a consequence of changes in retinal image and neural activity is also actively suppressed during saccades independent of visual input \cite{bremmer2009, binda2018}.
    
    \subsection{Saccade landing position prediction}
    
   		One of the earliest works on predicting the saccade landing position is that of Anliker~\shortcite{anliker1976}. Anliker's prediction method is based on Yarbus' observation~\shortcite{yarbus1965} that the saccade velocity profiles are approximately symmetric around the time when the peak velocity is reached and they estimate the landing position by doubling the displacement observed up to that point. Recently, there have been other studies based on the symmetry assumption \cite{paeye2016}. However, the saccade displacement profiles tend to get skewed for larger amplitudes such that the peak velocity is reached earlier than the midpoint of the saccade \cite{van1987}. Therefore, the prediction methods assuming a symmetric saccade velocity profile usually make an accurate prediction only for smaller saccades where the skewness is not very prominent in the velocity profile.
   		
   		To study the behavior of saccades, models based on Kalman-filter and higher order differential equations are introduced \cite{komogortsev2008, komogortsev2009, zhou2009}. While those models can be utilized for procedural simulation of eye movements, they require estimation of a large number of parameters. This could be time-consuming and inconvenient for predicting the landing position in real-time gaze contingent applications. As a more practical solution, Han et al.~\shortcite{han2013} introduced a compressed exponential model, while Wang et al.~\shortcite{wang2017} used Taylor series with a limited number of parameters to describe saccadic trajectory and make predictions for a short time window ($\approx\mkern-5mu10\ms$).
   		
   		\revcorr{Most gaze-contingent applications such as foveated rendering require accurate eye tracking and a maximum system latency around 50--70\ms. \cite{Albert2017}.} In order to combat system latencies typically observed in gaze-contingent rendering systems, Arabadzhiyska et al.~\shortcite{Arabadzhiyska2017} introduced a landing position prediction model based on the pre-programmed behavior of the saccades and the similarity of displacement profiles for similar saccadic amplitudes. For the same purpose, Griffith et al.~\shortcite{griffith2018,griffith2020a} proposed the use of support vector machine regression models and showed an extension to oblique saccades. Later, Morales et al.~\shortcite{morales2018, morales2021} proposed the use of Long Short-Term Memory (LSTM) networks for saccadic landing position prediction and Griffith et al.~\shortcite{griffith2020b} introduced a technique to improve the performance of LSTM and feed-forward network based models. Despite these active research efforts in saccade prediction for gaze-contingent rendering, investigation of different factors and their influence on the characteristics of saccades remain an open problem.

\section{Overview}
This work consists of two parts. In the first one, we present a user experiment (Section~\ref{sec:experiment}) where saccade profiles are collected for different amplitudes, orientations, depth levels, and with and without initial speed. In Section ~\ref{sec:experiment_results}, we analyze the collected data to discover the most significant factors affecting saccades. In the second part of this paper (Section~\ref{sec:method}), we present a method for adjusting existing saccade landing position prediction models to take the analyzed effects into account. 

\section{Experiment design}
\label{sec:experiment}
In our experiment, we aimed to investigate how saccade profiles depend on the saccade's orientation (in 3D space) and initial smooth pursuit eye movements. Additionally, we compared the effects with variability across different users. To this end, instead of conducting \revcorr{separate} experiments, each designed to investigate a single factor, we designed the stimuli and the task to simultaneously study all of the effects in different trials of a single experiment. As our main focus are applications of the saccade prediction techniques for head-mounted displays, the experiment was designed for a virtual reality device equipped with an eye tracker.

    \myfigure{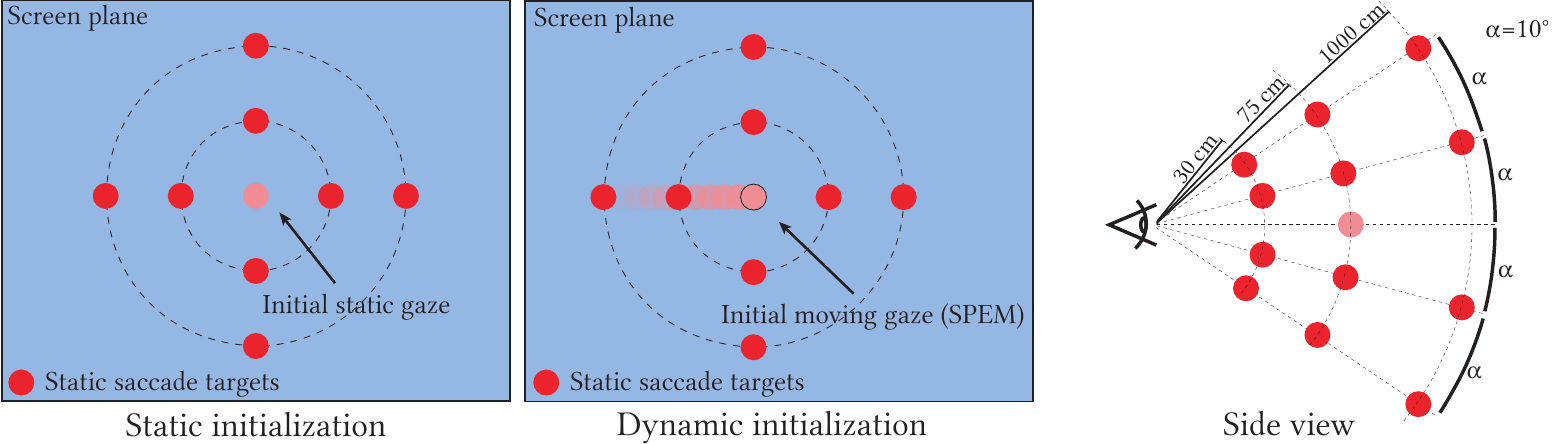} {\revcorr{The figure presents the main stages of each trial of our experiment. In the initial phase, we had either static initialization (left), where the initial gaze was shown as a static target in the center of the screen, or dynamic initialization (middle), where the initial gaze was moving to stimulate a smooth pursuit eye movement. After 1--2 seconds of the initial phase, the sphere was displaced to stimulate a saccade. Some trials of the experiment included a change in depth to stimulate vergence eye movement as shown on the right.}}
    
    
   \subsection{Stimuli}
   
\begin{wrapfigure}{R}{0pt}
        \includegraphics{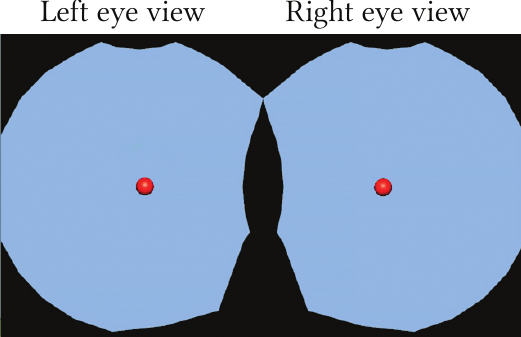}
        \caption{The images of the stimuli as shown to the participants of the experiment on the VR display.}
        \label{fig:VR_view}
    \end{wrapfigure}

   To guide the eye movements of participants, we rendered a red sphere on a blue background as the visual target (\refFig{VR_view}) at $75 \cm$ distance from the \revcorr{virtual camera}. In order to preserve the retinal size of the target as 1 visual degree throughout the experiment, the size of the rendered sphere was adjusted depending on its position and distance in 3D space. This prevented the potential saccadic inaccuracies during the experiment due to the changes in target size when the target displacement involved a change in depth (e.g., fixating on arbitrary parts of the target sphere when it appears bigger at a close distance). \revcorr{Each trial began with an initial phase where the participants are asked to either fixate on a \emph{static target} or follow a \emph{dynamic target}. The duration of the initial phase was randomly selected between 1--2 seconds to avoid anticipation effects.}
   
    \paragraph{Static initialization} \revcorr{In two-thirds of the trials, a static target appeared at the center of the screen, followed by a target displacement in one of the left, right, up, or down directions to stimulate a saccade between two static positions  (\refFig{figures/stimuli.pdf} - left). The displacement amounted to $10 \degree$ and $20 \degree$, respectively for short and long saccades. The trials with a change in vergence involved a simultaneous change in the depth with displacement (to $30 \cm$ or $1000 \cm$ w.r.t. virtual camera). The target remained visible for $2$ seconds at the end of each trial for fully completing the eye movement.}
    
    \paragraph{Dynamic initialization} \revcorr{In the remaining one-thirds of the trials, the target moved along a linear, vertical or horizontal trajectory with a constant velocity of $10 \dps$ (motion ramp) to stimulate smooth pursuit eye motion. Motion was followed by target displacement (step) to stimulate a saccade during smooth pursuit eye movement (a.k.a. \emph{ramp-step} paradigm,  \refFig{figures/stimuli.pdf} - middle). Target motion started from a source position located on the left/right or above/below the center of the screen for horizontal and vertical trajectories, respectively. The motion was always directed towards the center and it would last for a random duration of 1--2 seconds with the target never exceeding a distance of $10\degree$ from the center. Displacement in target step shared similar properties as the trials with a \emph{static target} (i.e., $10 \degree$ and $20 \degree$ displacement size with a single final depth of $75 \cm$ relative to the virtual camera position).}   
    
    
   
  
   \subsection{Task}
   During the experiment, each participant was asked to \revcorr{fixate on} or follow the target with their \revcorr{eyes}. The participants could abort the experiment at any time, especially if they started experiencing viewing discomfort. \revcorr{However, no participant has terminated the experiment prematurely due to viewing discomfort.}
   Each participant was shown the same set of stimuli, but in a \revcorr{randomized order to minimize the bias due to learning effect}. The set was constructed according to the cases visualized in \refFig{figures/stimuli.pdf} and contained combinations of: 
   \begin{itemize}
        \item 2 orientations of the saccade (horizontal and vertical), 
        \item 2 saccadic amplitudes ($10\degree$ and $20\degree$), 
        \item 3 depth levels to which the saccade was performed ($30 \cm$, $75 \cm$, or $1000 \cm$), and
        \item with/without initial SPEM.
   \end{itemize}
   To keep the experiment procedure simple for the participants, we excluded from our trials the cases where the sphere is moving in the initial phase and is then re-positioned to a different depth. We collected 12 saccades for each of the remaining cases amounting to 384 saccades per participant. The experiment took around 30 minutes to complete. To avoid fatigue, we divided the experiment into 3 sessions with 2 mandatory breaks of at least 10 minutes in between. We had 7 participants (2 of which are authors) with normal or corrected-to-normal vision, \revcorr{ages 25--37, all male}. Due to amplified eye tracking inaccuracies associated with the use of eye glasses during a pilot run of our experiments, the participants with corrected-to-normal vision only used contact lenses. \revcorr{Also, to avoid calibration related problems, we introduced an additional verification step after the eye tracker calibration: The users were asked to consecutively fixate on four different targets, also red spheres with visual size of $1\degree$, located at $10\degree$ in the periphery in the four primary directions. We repeated the calibration procedure if the estimated gaze location was more than $1\degree$ away from any of the four targets.}

    \subsection{Hardware}
    The experiment was implemented using Unity\footnote{https://unity.com} platform and it was ran on HTC Vive Eye Pro headset which provides 1440$\times$1600\,px resolution per eye at 90\Hz. We used the headset's integrated 120\Hz\, eye tracker which was calibrated at the beginning of each session using the 5-point calibration procedure provided by the eye tracker software. \revcorr{The accuracy of the eye tracker reported by the manufacturer is $0.5\degree$--$1.1\degree$, however, recent research \cite{Sipatchin2021} reports different values: $4.16\degree$ mean average accuracy of both eyes across field of view of $27\degree$ and mean precision of $2.17\degree$ for a head-still condition such as our task; the data loss is estimated to be $3.69\%$. }

\section{Analysis of experimental data}
\label{sec:experiment_results}
The data from the experiments was used to extract mean saccade profiles which were then analyzed to quantify the influence of different factors. To our knowledge, there is not any common dissimilarity measure for comparing saccadic displacement profiles with each other. Therefore, we also provide a formulation of our measure that helps detecting the most significant factors affecting the saccade.

\subsection{Saccade profiles extraction}
\label{sec:saccade_profiles_extraction}
Similar to Arabadzhiyska et al.~\shortcite{Arabadzhiyska2017}, our saccade profiles describe the on-screen displacement with respect to the saccade anchor point, as a function of time that elapsed since the beginning of the saccade. \revcorr{To extract the profiles from the data collected in the experiment, we follow the procedure described in \cite{Arabadzhiyska2017}. We first use a high velocity threshold value for detecting a saccade and then a second, lower velocity threshold value to scan the gaze samples backward in time to find its beginning. The first step gives us the detection point of the saccade and the second - its anchor point at which we assume the saccade has started. This two-step procedure reduces the detection likelihood of false positives and collects the additional samples that are needed to capture the beginning of the saccade. Since we compute the velocity by estimating the distance of consecutive samples without applying a velocity filter, a double threshold policy improves the reliability of correctly detecting saccades. For further details, please refer to the original paper.}

\begin{wraptable}{R}{0.35\textwidth}
    \vspace{0.5em}
	\caption{\revcorr{The factors that we consider when analyzing saccades and the categories in which we classify them according to each individual factor. }}
	\label{tbl:groups}
	\begin{center}
	\begin{tabular}{@{}ll@{}}
		\toprule
		Factors                  & Categories  \\ \midrule
		\multirow{2}{*}{{\sc Orientation}} & {\sc horizontal}     \\
		& {\sc vertical}  \\ \midrule
		\multirow{3}{*}{{\sc Depth}} & {\sc same} \\
		 & {\sc nearer} \\
		 & {\sc farther}  \\\midrule
		\multirow{3}{*}{{\sc Initial movement}} & {\sc static} \\
		 & {\sc same}     \\
		 & {\sc opposite} \\\midrule
		{\sc Users} & \textit{Each user} \\ \midrule
		{\sc Amplitude} & {\sc $-1\degree$}, {\sc $+1\degree$} \\
		\bottomrule
	\end{tabular}
	\end{center}
\end{wraptable}

\revcorr{To analyze the effects of different factors, we define sets of categories belonging to each factor and we classify each saccade of our dataset into one of its categories. Each category contains a subset of the dataset and within the same factor the categories are mutually exclusive. }
To investigate the influence of orientation of the saccade, we classify the saccades according to the location of their landing position with respect to the initial position of the gaze (\revcorr{factor}: {\sc Orientations}, \revcorr{categories}: {\sc horizontal}, {\sc vertical}). Similarly, to analyze the influence of depth/vergence change, we classify saccades according to the depth of final position with respect to the initial point (\revcorr{factor}: {\sc Depth}, \revcorr{categories}: {\sc same}, {\sc nearer}, {\sc farther}). For analyzing the influence of SPEM, we classify the initial eye motion at the beginning of the saccade which may be performed from a static target, a target moving in the direction of the imminent saccade, and a target moving in the opposite direction of the imminent saccade (\revcorr{factor}: {\sc Initial movement}, \revcorr{categories}: {\sc static}, {\sc same}, {\sc opposite}). \revcorr{Additionally, to analyze differences among subjects we create a category for each person containing only the saccades performed by this individual.} (\revcorr{factor}: {\sc Users}, \revcorr{categories}: \textit{each user}). We analyze the factors for short (amplitude $10 \degree$) and long (amplitude $20 \degree$) saccades separately to verify that observed effects are consistent across amplitudes. 

In gaze-contingent rendering applications, inaccuracies in saccade prediction may remain imperceptible if the prediction error is limited. Previous research on the anatomy of the human retina revealed that the angular subtense of the human fovea is approximately $4\degree$--$5\degree$ \cite{hendrickson2005}. We assume that when the prediction error reaches around approximately half of this distance, the misplacement of foveal region becomes visible by observers. Consequently, an improvement of the prediction error may be evaluated by comparing with this baseline. Therefore, we also included the saccades at a range of $2\degree$ difference in amplitude around short and long saccades. More specifically, we consider saccades with amplitudes $9 \degree$ and $11 \degree$ for the short saccades, and $19 \degree$ and $21 \degree$ for the long saccades in our comparisons (\revcorr{factor: {\sc Amplitude}, categories}: $-1\degree$, $+1\degree$). A summary of the factors and the categories we defined for our experiments are shown in Table~\ref{tbl:groups}.

To analyze the differences within each \revcorr{factor}, we aim to compute mean profiles for each \revcorr{category created for it} and for each saccade amplitude $\alpha \in \{10 \degree$, $20 \degree\}$. For each category, we start by filtering out the saccades that do not belong to it and then align the displacement profiles in the temporal domain. For each \revcorr{factor} and each amplitude $\alpha$, we start by removing all the saccades with amplitude outside the range $[\alpha - 1\degree, \alpha + 1\degree]$. In addition, we check the length of the saccades, the direction of SPEM, and the direction of the vergence performed by the participants to label the samples that do not conform with the expected behavior in the category as outliers. Then we align the anchor points of saccades by applying a temporal offset \cite{Arabadzhiyska2017}. In our case, we choose the velocity threshold for detecting a saccade as  $180\dps$ and the anchor point as $90 \dps$, \revcorr{the same values used by Arabadzhiyska et al.~\shortcite{Arabadzhiyska2017} for their subjective experiment}. All eye-tracker samples $30\,ms$ prior to the anchor point are included in the analysis as well. To obtain the mean profile sampled at equal time intervals, we resample each profile using linear interpolation of measured displacements. \revcorr{Our eye tracker operates at $120\Hz$ and provides a gaze estimations every $8\,ms$ or $9\,ms$ and not all of these samples are valid. Therefore, the samples for the different saccades are at different time positions with respect to the their beginnings. Resampling at equal intervals is needed to align the displacement values for each saccade in the same time positions. We chose our interval to be $1\,ms$. } Since the endpoint of each saccade occurs at an arbitrary time, we consider the endpoint of the mean profile to be positioned at the mean time position of all the endpoints.

After the above initial processing, the mean profiles are computed by averaging samples of all saccades within each category. Formally, we represent those mean profiles as a sequence of $N$ mean samples computed from the original profiles:
\begin{equation} \label{eq:s_k}
    \mean{S} = \{\mean{s_{0}}, \mean{s_{1}}, ..., \mean{s_{N}} \},
\end{equation}
where each sample $\mean{s_{l}} = (t_{l}, \mean{d_{l}}, \sigma_{l})$ is defined by its timestamp $t_{l}$, mean displacement $\mean{d_{l}}$, and the standard deviation of all the displacement values for the given timestamp $\sigma_{l}$ within the category. The first sample of the saccade ($\mean{s_{0}}$) is the anchor point ($t_{0} = 0$) whereas the last sample ($\mean{s_{N}}$) is the end point ($\mean{d_{N}}$) and it is equal to the amplitude of the saccade. \revcorr{ \refFig{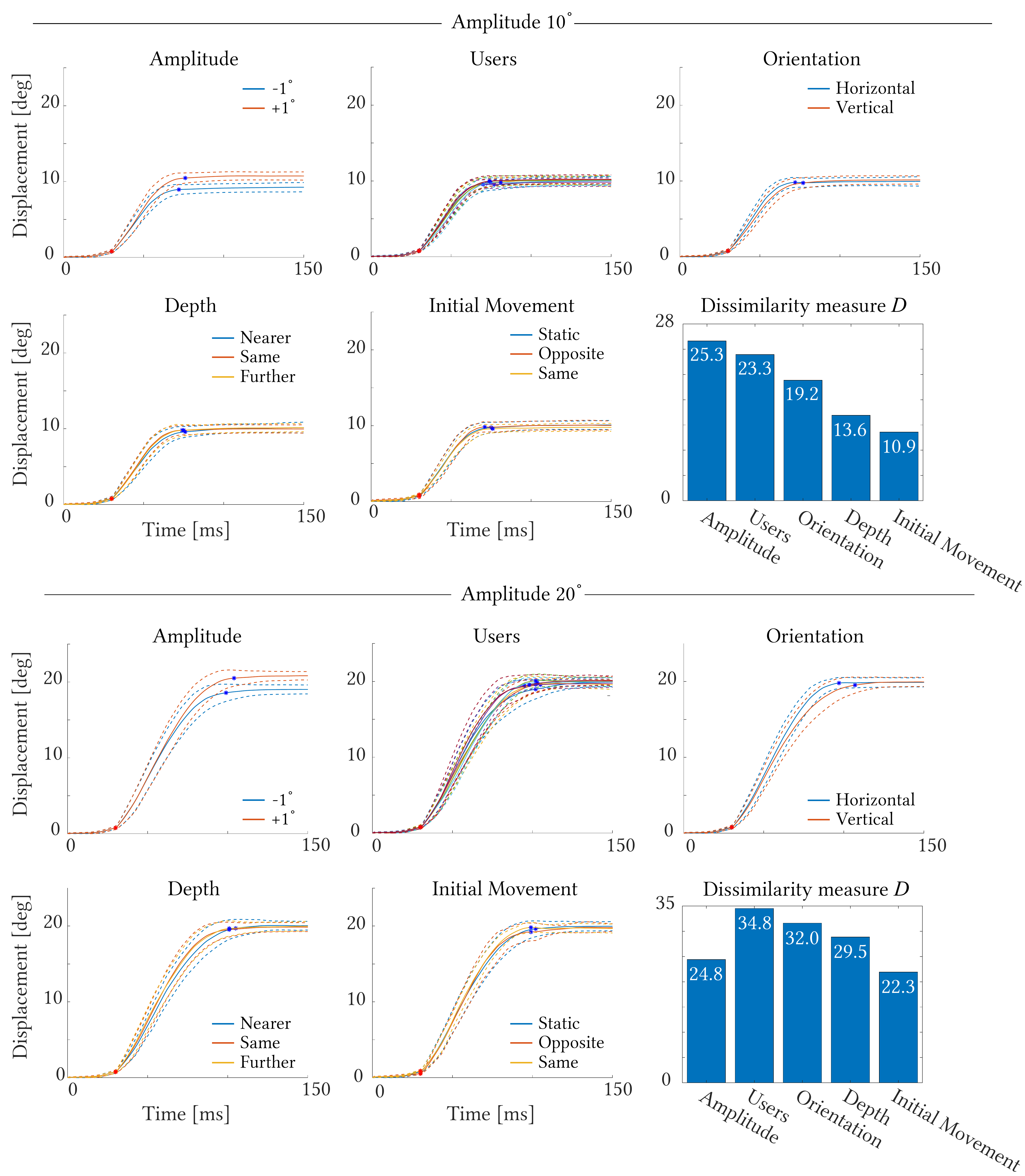} visualizes the mean displacement profiles for all categories grouped by the factors they belong to.} 

\subsection{Dissimilarity measure for saccade displacement profiles}

\noindent\begin{minipage}{.65\textwidth}
\begin{minipage}{0.95\textwidth}
\revcorr{To be able to analyze and compare the effects of different factors, we propose a dissimilarity measure for quantifying the differences between the mean saccade profiles corresponding to individual categories within a single factor}. More precisely, for a given set of mean profiles $\{\mean{S^k} | \mean{S^k} = \{\mean{s^k_0}, \mean{s^k_1}, ..., \mean{s^k_{N_k}} \}\}$ \revcorr{belonging to the categories within a factor (Table~\ref{tbl:groups}), we define a measure that correlates with the differences for that factor as}:
\vspace{1ex}
\begin{equation}
\label{eq:similarity}
    D(\{\mean{S^k}\}) = \sum_{l = 0}^{\min\limits_{k} N_k}  \dfrac{\max\limits_{k} \mean{d^k_l} - \min\limits_{k} \mean{d^k_l}}{\max\limits_{k} \sigma^k_l},
\end{equation}
\vspace{1ex}
where $k$ is the index of a category. The measure can be seen as an area between the upper and lower envelope of all mean displacement profiles for the factor ($\{\mean{S^k}\}$), normalized by the maximum standard deviation of the displacement values observed for the factor ($\max\limits_{k} \sigma^k_l$). 
\revcorr{\refFig{dissimilarity_measure} illustrates an abstract example of the mean displacements ($\mean{d^k_l}$) and the standard deviations ($\sigma^k_l$) of three hypothetical sample categories.}
It is important to note that this measure requires the mean displacement profiles to have equal sampling intervals. 
\hspace{1.5ex}
\end{minipage}
\end{minipage}
\begin{minipage}[r]{.35\textwidth}
\centering
  \includegraphics[width=\textwidth]{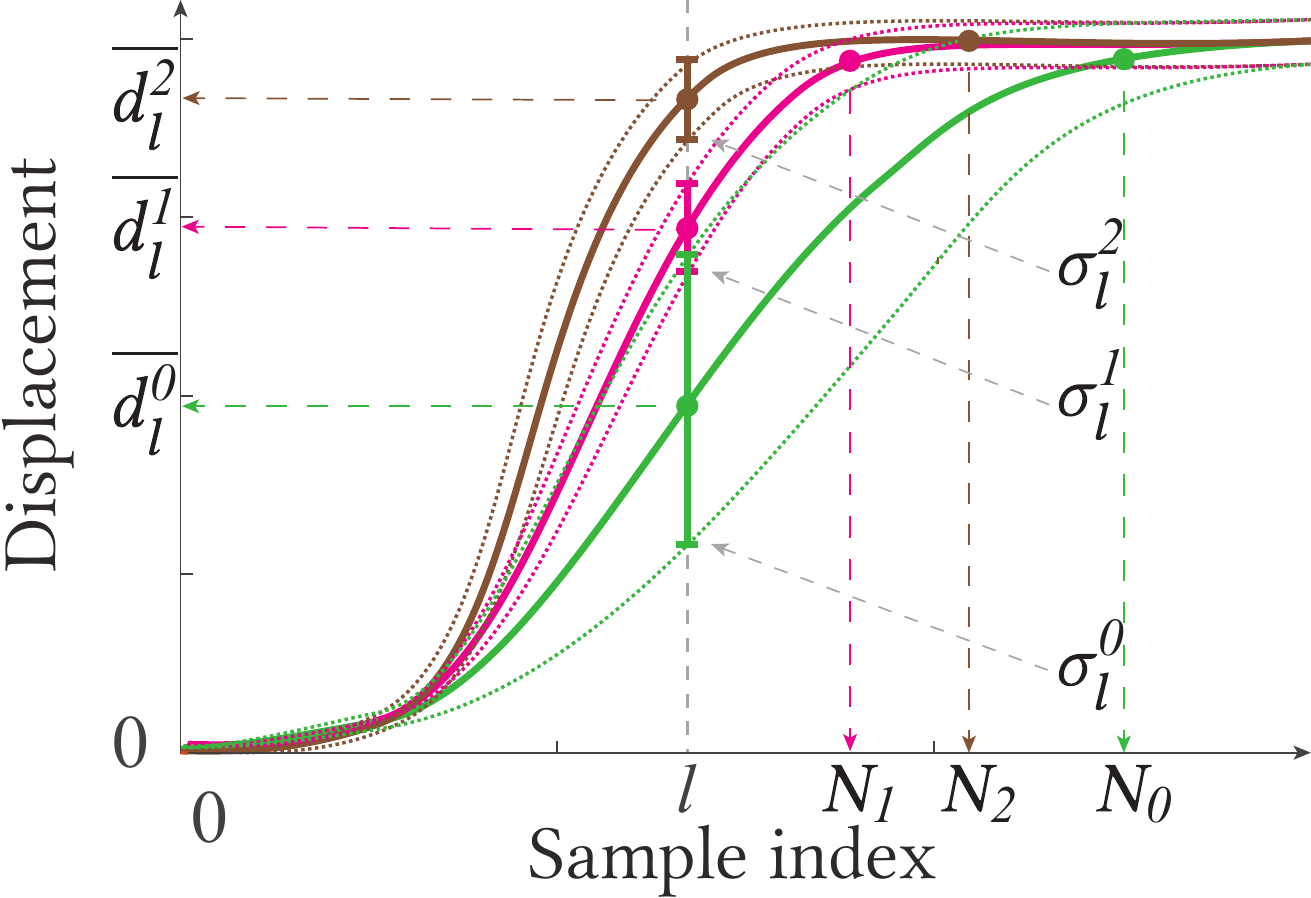}
  \captionof{figure}{Mean displacements ($\mean{d^k_l}$) and standard deviations ($\sigma^k_l$) that we used in \refEq{similarity} are shown for hypothetical mean saccade profiles. Mean displacement profiles of three categories; namely, $\mean{S^0}$, $\mean{S^1}$, and $\mean{S^2}$, are represented by green, pink, and brown solid lines, respectively, whereas whiskers and dotted lines visualize the standard deviation of saccade displacements from the corresponding category.}
  \label{fig:dissimilarity_measure}
\end{minipage}
\vspace{0.1ex}

Computing the dissimilarity measure in \refEq{similarity} yields a higher value \revcorr{if there is a more significant difference between the mean saccade displacement profiles corresponding to different categories within a given factor (\refTbl{groups}). The differences between the categories} commonly manifest themselves through speed ups or slow downs in saccade displacement profiles and we use our dissimilarity measure for identifying perceptually significant changes to the displacement profiles that require an update to the prediction model or training data to avoid visual artifacts.
  
\myfigurelong{figures/users_4_amp10_and_20_summary.pdf} {Effects of different factors on the saccade mean displacement profiles computed from our experiment data. The solid lines represent mean profiles for each category, while the dashed lines visualize corresponding standard deviations. The bar plots show the values of our profile dissimilarity measure (\refEq{similarity}) for different factors (\refSec{saccade_profiles_extraction}). The bars representing the {\sc Amplitude} factor are provided as a reference baseline for the minimum value of similarity to observe a significant effect (please refer to \refSec{saccade_profiles_extraction} for details).}
  
\subsection{Discussion}
\label{sec:experiment_discussion}
    \refFig{figures/users_4_amp10_and_20_summary.pdf} summarizes the effects that different factors have on the mean displacement profiles. Additionally, we provide a bar plot of our dissimilarity measure for each factor.
    
    A clear difference can be observed for the case where we compare mean displacement profiles with different amplitudes (factor: {\sc Amplitude}). The difference for both $10 \degree$ and $20\degree$ saccades is consistent with the fact that longer saccades exhibit steeper ascend in their displacement profiles compared to shorter saccades. Existing saccade landing prediction models depend on these profiles to be distinguishable, which is an expected effect. It also serves as the baseline to compare the difference exhibited by the other factors of interest as we mentioned in Section~\ref{sec:saccade_profiles_extraction}. Therefore, we aim to identify the effects that will change the performance of saccade landing position prediction and assume that factors that provide smaller effects than what is observed at $2\degree$ change in the saccade amplitude may not lead to significant improvements in applications that rely on the prediction. In particular, the value of the dissimilarity measure $D$, $25.3$ for $10 \degree$, and $24.8$ for $20 \degree$ saccade are the reference points for analyzing the effects of the other factors.
    
    \begin{wrapfigure}{R}{0.5\textwidth}
        \centering
        \includegraphics[width=0.5\textwidth]{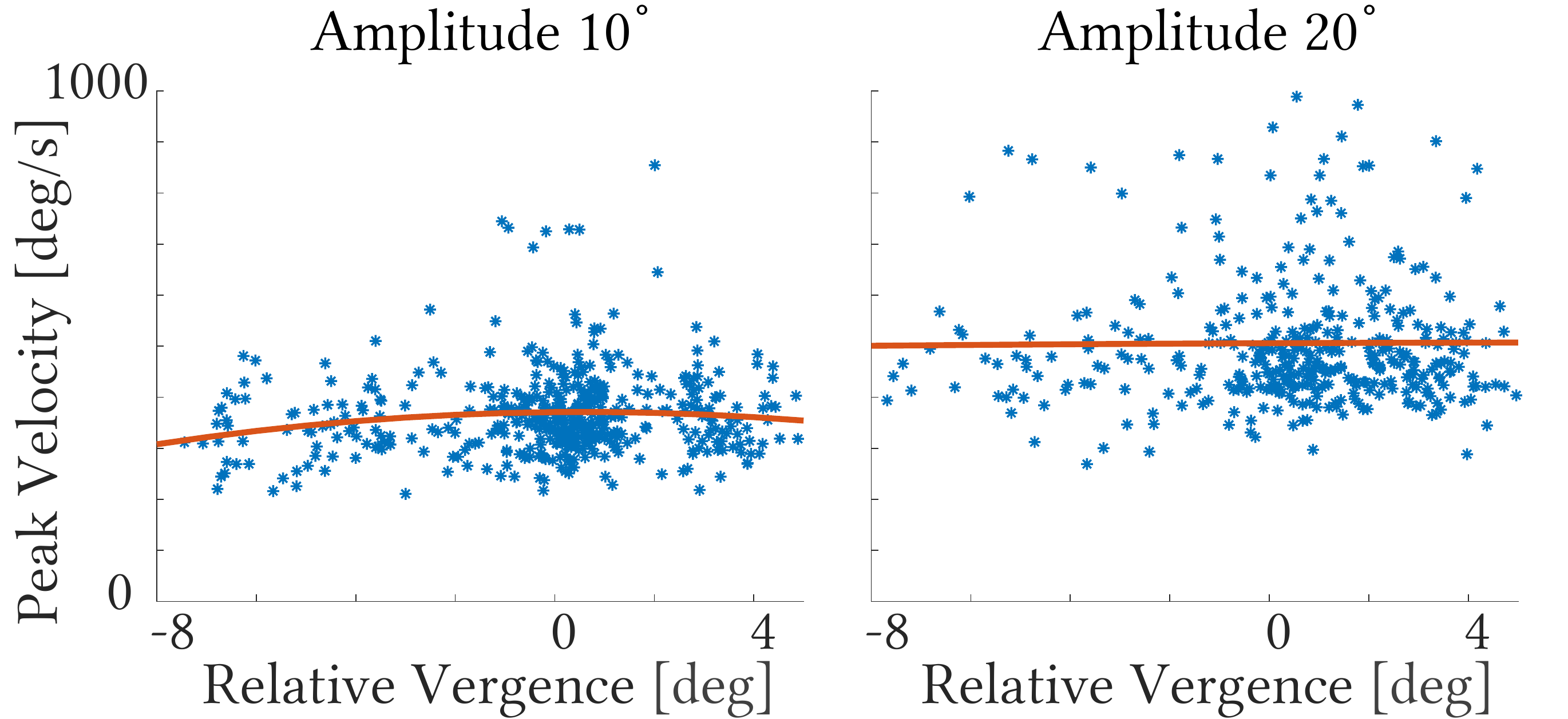}
        \caption{The relation between the vergence change during the saccade and the peak velocity. The blue points correspond to individual saccades, while red curves are the quadratic line fits showing the overall trend. \revcorr{Negative values indicate saccades that move closer to the observer, whereas positive values indicate saccades moving further away. }}
        \label{fig:vergence}
    \end{wrapfigure}
    
    Apart from the {\sc Amplitude} factor, the most significant differences were observed for {\sc Users}. The differences become even more apparent for longer saccades ($20 \degree$). It has been already shown by Arabadzhiyska et al.~\shortcite{Arabadzhiyska2017} that tailoring a model to fit the personal saccadic characteristics of a user leads to lower saccade prediction error and to a higher subjective preference for that user compared to the model trained for the average population. While they demonstrated this in a task-performance experiment, here we demonstrate the underlying difference in saccade profiles.
    
    The third factor with the highest differences was {\sc Orientation}. Similar to {\sc Users}, the differences for $10 \degree$ saccades were smaller than for the {\sc Amplitude} factor, but the opposite can be observed for $20 \degree$ saccades. For this factor, the differences become close to those observed with the {\sc Amplitude} factor.
    
    Contrary to our expectations, moving the target to different depth levels (factor {\sc Depth}) led to smaller changes in the mean displacement profiles, especially for $10 \degree$ saccades. While we observed some changes in the peak velocity (\refFig{vergence}), the differences are smaller than those reported by the previous studies (Section~\ref{sec:saccade_characteristics}). We relate this discrepancy with the existing studies mainly to the profound difference between the real and virtual environment. First, standard head-mounted displays are not fully capable of reproducing accommodative cues, and any depth change only results in a change in the vergence (due to the change in disparity), but it does not trigger an accommodation response from the participants' visual system. The lack of an accommodation response may be seen as a deviation from real-world viewing conditions, but it applies to most mainstream stereoscopic HMDs used for virtual reality. Therefore, we have not tried to mitigate this effect in our experiments. Second, similar to all experiments conducted on stereoscopic displays with a lack of accommodation response, the presence of well-known vengeance-accommodation conflict \cite{Shibata2011} imposed a limit on the depth ranges that we could test in our experiments without causing viewing discomfort for the participants. These differences between virtual reality and real-world viewing conditions may explain the discrepancy between our measurements and the previous studies, most of which are conducted under real-world viewing conditions. \revcorr{Additionally, the choice of the stimuli could affect the outcome of our experiments. While the small spheres used in the experiment enable precise control over the participant's gaze location and saccades, the fact that they do not change their size according to the distance removes the size cue. The lack of this cue could potentially influence the saccade accuracy. Also, the use of specific colors, red and blue in our case, may lead to a different amount of edge blur due to the wavelength-dependent accommodation.}

    With the final factor, {\sc Initial movement}, we observed that the initially moving target led to smaller differences in the displacement profiles. We believe that higher pursuit speeds could potentially enhance the effect. In our experiment, we chose a moderate pursuit speed (10\dps) to keep the task simple and give the observer ample time to properly fixate on the moving target and initiate SPEM. Similar to {\sc Depth}, the differences become larger for $20 \degree$ saccades, and they are close to these observed with the {\sc Amplitude} factor. It is possible that the differences become more apparent for more extreme saccade amplitudes. Unfortunately, reliable measurement of larger amplitude saccades poses problems due to the fact that virtual reality headsets have a limited field of view with high fidelity.

    \revcorr{In all our experiments, we used an optical eye-tracker, the current technology of choice for VR and AR applications. Despite its widespread use, this technology is not suitable for capturing all the characteristics of eye movements \cite{Nystrom2013, Hooge2015, Nystrom2016, Hooge2016}. In particular, due to post-saccadic oscillations of the pupil, the optical eye trackers have low accuracy in estimating the saccade onset, peak velocity, and its end. Additionally, the sensitivity of the eye trackers to the changes in the pupil's size \cite{Drewes2014, Jaschinski2016, Hooge2019} has a detrimental effect on the correct estimation of the vergence and the binocular fixation point. To address these limitations and measure eye movements more accurately it is possible to use eye-tracking technology such as the wearable scleral coil tracking system proposed by Whitmire et al.~\shortcite{Whitmire2016}. However, most users might find coils to be a very invasive way to track their gaze orientation, and to our knowledge, no commercial VR or AR headset uses such technology. Therefore, in our work, we focus on optical eye tracker technology, which, despite its limitations, has been already shown to be beneficial in applications such as foveated rendering \cite{guenter2012,patney2016,Arabadzhiyska2017}. At the same time, it is important to note that the generalization of our findings to of scleral coil eye-tracking technology needs further investigation.}

    In the remaining part of the paper, we demonstrate a new technique that accounts for differences in the saccade profiles to provide better saccade prediction. For demonstration purposes, we chose to focus on two factors that exhibit the highest differences, i.e., {\sc Users} and {\sc Orientation}. While the personalization of the prediction model for a specific user was demonstrated in \cite{Arabadzhiyska2017}, the process required collecting a large set of saccades. Here, our goal is to reduce the amount of required data. On the other hand, to our knowledge, adjusting existing models to adapt to the saccade's orientation has not been done before, but our findings suggest that it could improve the prediction accuracy. Therefore, designing prediction methods or adjusting the existing ones to handle different orientations correctly may provide additional benefits in the final applications.

\section{Method for tuning saccade prediction models}
    \label{sec:method}
    In Section \ref{sec:experiment_results}, we analyzed how different factors affect the displacement profiles of the saccades. We observed the dissimilarity for the individual factors to be comparable to the dissimilarity for the {\sc Amplitude} factor, with the biggest ones for {\sc Users} and {\sc Orientation} factors. The observed dissimilarities suggests that incorporating factors such as saccade orientation or the difference among users may improve the saccade prediction. However, the fundamental problem in deriving a model which captures such dependencies lies in data collection. Individual saccades collected for training such models contain noise; therefore, many of them have to be combined to create a reliable prediction. For example, the prediction model proposed by Arabadzhiyska et al.~\shortcite{Arabadzhiyska2017} required each participant to perform 300 saccades. Still, the model does not capture factors other than the saccade amplitude. Consideration of additional factors, such as orientation, depth, and SPEM, would significantly increase the number of the required saccade samples, making the data collection for individual users tedious and sometimes infeasible. Similarly, most machine-learning approaches, such as Morales et al.~\shortcite{morales2018}, have high data demands for training.
    
    To address the problem of data collection, we propose an alternative approach. Instead of exhaustively collecting data from psychophysical experiments, which enables training prediction models to capture all factors, we postulate that the influence of many factors, such as orientation or user, can be \revcorr{approximated} by a low-parameter transformation of the data. The advantage of such a solution is that the effect of additional factors is captured using a small number of parameters, and therefore, such a model is more robust to noise and the reduced number of collected saccades. Successful applications of this approach are shown in the past, such as the method of Lesmes et al.~\shortcite{Lesmes2010}, which uses the a priori information about the contrast sensitivity function's (CSF) general functional form to maximize the information gained from a small number of measurements. Similarly, in this work, we seek a global transformation of the profiles of a saccade prediction model, which has a small number of parameters, yet allows for explaining the effects of additional factors influencing the saccade performance.
    
    The main observation behind our solution is that the differences in the saccade profiles can be attributed to the changes in the saccades' performance/velocity caused by the factors that we investigated in our experiments. This observation can be made by looking at the differences among slopes of the individual mean saccades profiles in \refFig{figures/users_4_amp10_and_20_summary.pdf}. We demonstrate that these changes can be effectively modeled by shearing the profiles parallel to axis representing the time domain (\refFig{data_shift}). Additionally, we observe that the appropriate shearing factor changes with saccade's amplitude, but we show that this change can be approximated with a low-degree polynomial. This is the key to our technique, as it allows us to compute the shear factor for few saccade amplitudes and then interpolate or extrapolate the shearing transform to the other amplitudes. Below, we provide a formal definition of shearing (Section \ref{sec:shearing_profiles}) and a shear between two saccade profiles (Section \ref{sec:optimal_shear}). Then, we describe the derivation of the shearing-based transformation of saccade profiles and how it can be applied to modify a prediction model to account for additional factors in Section \ref{sec:application}.   
   
    \subsection{Shearing saccade profiles}
    \label{sec:shearing_profiles}
    Given a saccade profile $S = \{s_0, s_1, ..., s_N \}$, where each sample is defined by a couple of scalars $s_l = (t_l,d_l)$ representing time stamp, $t_l$, and corresponding displacement, $d_l$, we define a sheared version of the profile by applying a 2D shearing parallel to the time axis followed by resampling to restore uniform sampling in time domain. More formally, to shear the profile $S$ with a shearing factor $\lambda$, we first transform its samples using a 2D shearing matrix:
          \begin{equation}
            \begin{bmatrix}
                \widehat{t_l}\\
                \widehat{d_l}
            \end{bmatrix} = 
            \begin{bmatrix}
                1 & \lambda\\
                0 & 1
            \end{bmatrix}
            \begin{bmatrix}
                t_l\\
                d_l
            \end{bmatrix}, \quad \lambda \in [-1; 1].
        \end{equation}
    The resulting profile $\widehat{S} = \{(\widehat{t_1},\widehat{d_1}), (\widehat{t_2},\widehat{d_2}), ..., (\widehat{t_N},\widehat{d_N}) \}$ is not sampled regularly at 1\,\ms~intervals anymore after applying the shearing transformation because the time stamp, $t_l$, of each sample changes. Therefore, we apply a simple linear interpolation to resample it back to 1\ms~intervals and obtain the final sheared profile. In the rest of the paper, we denote shearing as a function $\Psi$, and a saccade or mean saccade profile $S$ sheared with shearing factor $\lambda$ as $\Psi(S,\lambda)$.
  
  \subsection{Computation of shearing transformation between saccadic profiles}
  \label{sec:optimal_shear}
\begin{figure}
    \centering
    \begin{minipage}{.66\textwidth}
      \centering
        \captionsetup{width=\textwidth}
        \includegraphics[width=\textwidth]{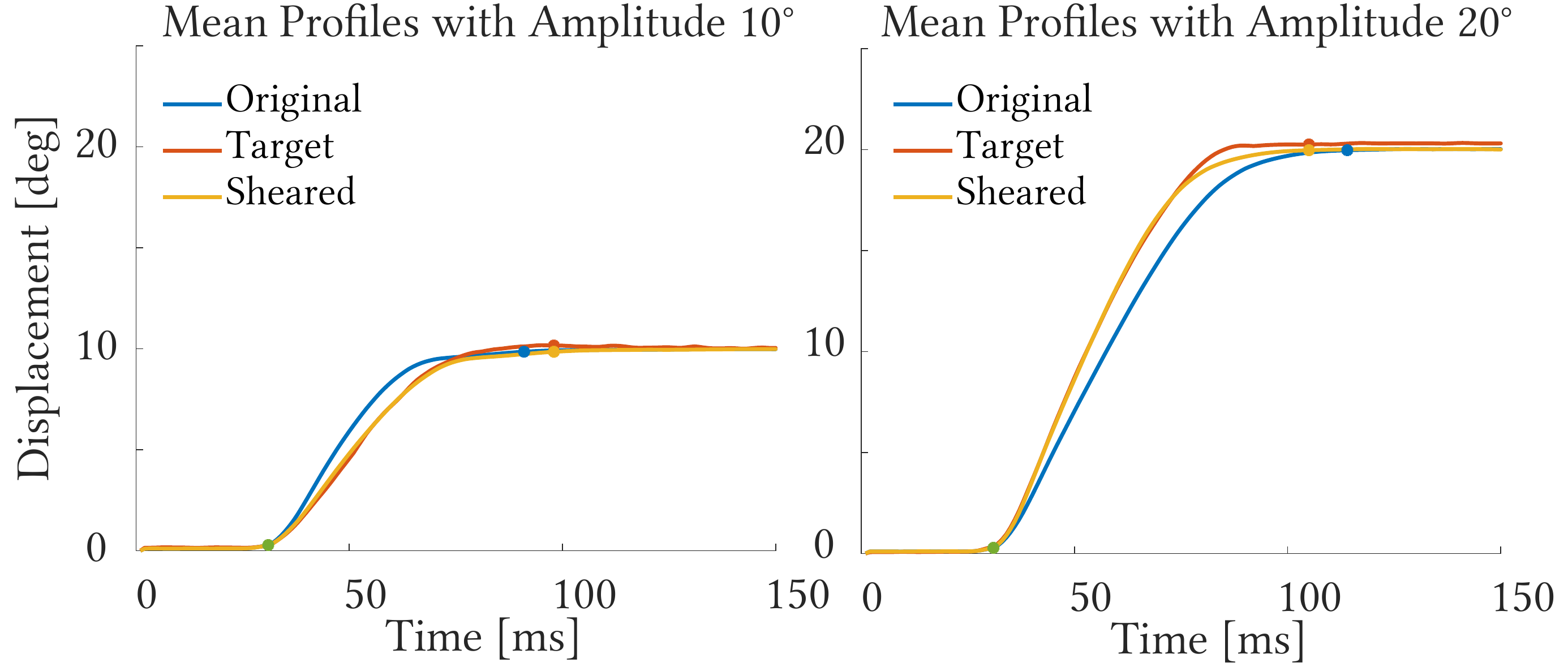}
        \caption{\revcorr{Two examples of shearing original mean saccade profiles to match different targets. The target on the left represents a category with slower saccades than the original. The target on the right represents a category with faster saccades.}}
        \label{fig:data_shift}
    \end{minipage}%
    \begin{minipage}{.33\textwidth}
      \centering
      \captionsetup{width=0.85\textwidth}
      \includegraphics[width=\textwidth]{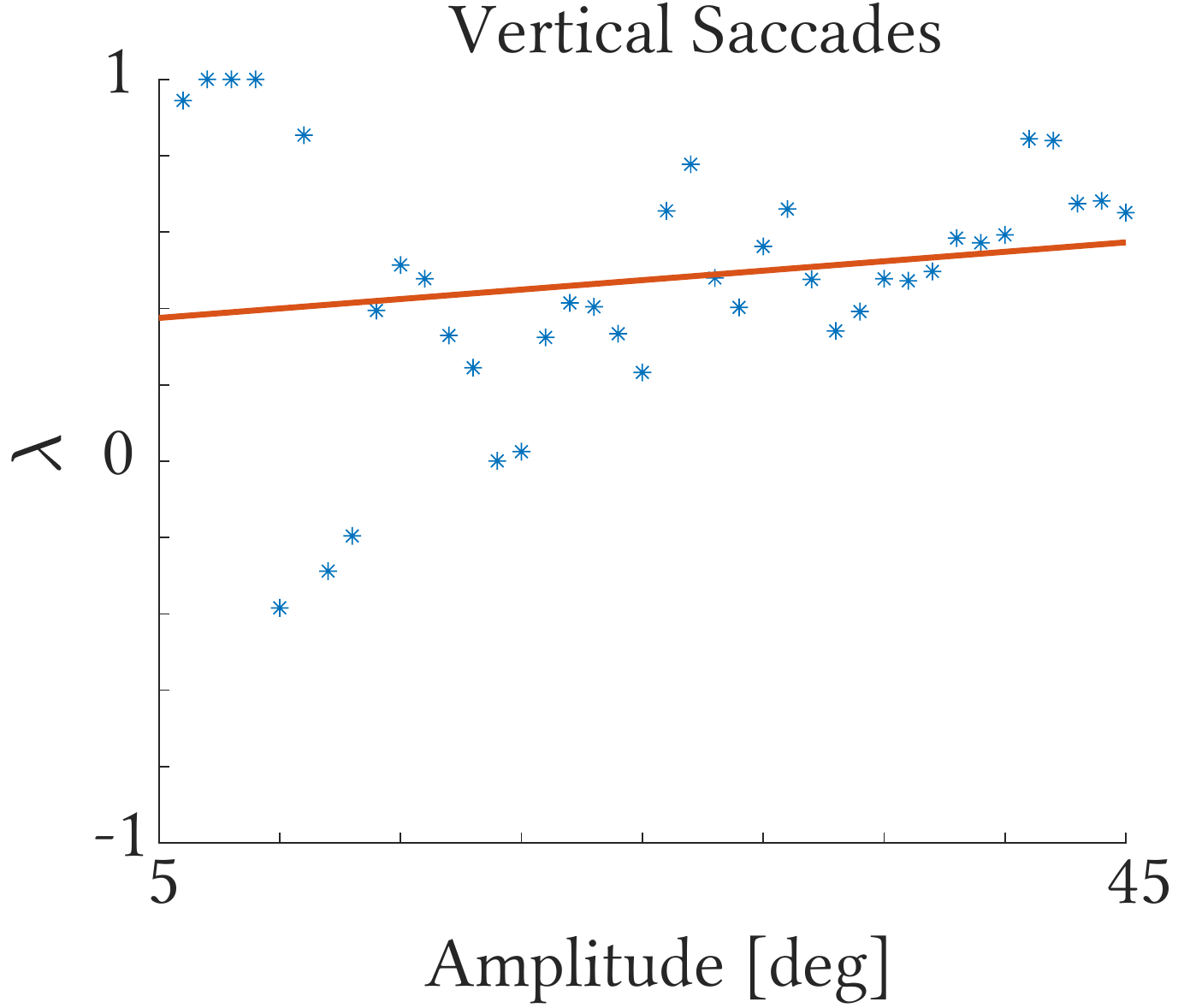}
      \caption{The linear function $f(\alpha)$ (red) is fitted to the shearing factors $\lambda_{\alpha_i}$ (blue).}
      \label{fig:lambda_fit}
      \end{minipage}
  \end{figure}

  Given two saccade profiles $S^k = \{s^k_0, s^k_1, ..., s^k_N \}$, where $k \in \{1,2\}$ and $s^k_l = (t^k_l,d^k_l)$, we can compute a shearing factor $\lambda$ that describes the difference between those two profiles. Formally, we define the shearing between $S^1$ (original profile) and $S^2$ (target profile) as $\lambda = \Lambda(S^1, S^2)$ for which the 2D shear applied to $S^1$ minimizes the difference with respect to $S^2$, i.e.,:
  \begin{equation}
  \Lambda(S^1, S^2) = \mathop{\mathrm{argmin}}_\lambda \sum_{l = 0}^N   |d^*_l - d^2_l|, \quad \text{subject to } S^* = \Psi(S^1,\lambda).
  \label{eq:Lambda}
  \end{equation}
  This definition relies on same sampling of time domain by all profiles involved in the computation ($S^1$, $S^2$, $S^*$). This is, however, guaranteed by the definition of $\Psi$. The above minimization problem can be easily solved using binary search. Figure~\ref{fig:data_shift} demonstrates two examples of how the shear between two saccade profiles can be used to align them.

\subsection{Application}
\label{sec:application}
Previous models for predicting saccades, such as \cite{Arabadzhiyska2017,morales2018,morales2021}, are trained on large datasets containing saccades with various amplitudes and orientations collected from multiple users using an eye tracker. These models do not account for all the factors analyzed in Section~\ref{sec:experiment_results}. Here, we demonstrate how to use the shearing strategy described in Section~\ref{sec:shearing_profiles} and \ref{sec:optimal_shear} to account for these factors. It is possible to apply the shearing transformation to saccade displacement profiles directly if the dataset is available (\textit{data shear}). In some cases, although the model is accessible, the dataset that the models were trained on may not be available. For such cases, if possible to extract saccade displacement profile approximations from the model, we apply the transformation to the recovered profiles instead (\textit{model shear}).

\paragraph{Data shear}
The first approach we consider that utilizes the idea of shearing the saccade profiles is to transform all the saccades in the dataset to create a new dataset that represents a particular type of saccades and then recompute the prediction model using the augmented dataset instead of the original one. In particular, we consider here shearing the saccade profiles to create a dataset and models for horizontal, vertical, and personalized saccades. In all three cases, we apply our shearing strategy in the same way. First, to obtain the specific saccades for each category (horizontal, vertical, or a particular user), we extract the corresponding saccades from the dataset. These saccades act as a target for the required shear computation applied to the remaining saccades to compute the final dataset. We then discretize the amplitude domain. In our experiments, we chose the discretization step to be $1\degree$. For each discrete amplitude value $\alpha$, we estimate the mean displacement profile by averaging the saccade displacement profiles with amplitudes in the range $\{\alpha - 1; \alpha + 1\}$ for both the original dataset and for the target dataset following the same procedure as described in Section~\ref{sec:saccade_profiles_extraction}. Here, we denote the mean profiles of amplitude $\alpha$ in the original dataset as $\mean{S^\alpha_o}$ and the target dataset as $\mean{S^\alpha_t}$. \revcorr{The number of saccades constructing $\mean{S^\alpha_t}$ we denote with $|S^\alpha_t|$}. Note that $\mean{S^\alpha_o}$ and $\mean{S^\alpha_t}$ are mean saccade profiles of the same amplitude. The only difference is that $\mean{S^\alpha_o}$ comes from the original dataset, which contains all types of saccades (e.g., all orientations) while $\mean{S^\alpha_t}$ is a mean profile for the specific category (e.g., horizontal, vertical, or for a particular user). The goal is to use this correspondence to define the shear that needs to be applied to the original dataset, to make it represent a particular category of saccades. To this end, we compute a series of shearing factors $\{\lambda_{\alpha_0}, \lambda_{\alpha_2}, ..., \lambda_{\alpha_M}\}$ for each amplitude $\alpha_i$ following \refEq{Lambda}, i.e., $\lambda_{\alpha_i} = \Lambda(\mean{S^{\alpha_i}_o}, \mean{S^{\alpha_i}_t})$.

The main objective of such a dataset derivation is to obtain a large dataset of saccades while using only few measured profiles. To this end, we propose to first collect a subset with a particular category of saccades and compute the shear (Section~\ref{sec:optimal_shear}) of the mean profiles with respect to the mean profiles in the large dataset. Using this procedure, we obtain the relationship between the profiles in the large dataset and newly collected one for a few saccade amplitudes. To obtain the shearing factors for the whole range of saccadic amplitudes, we use a linear regression to fit the linear function $f(\alpha)$ that minimizes:
\begin{equation}
    \revcorr{\sum_{i=1}^M \frac{\left|f(\alpha_i) -  \lambda_{\alpha_i}\right|}{|S^\alpha_t|},}
\end{equation}
\revcorr{where $|S^\alpha_t|$ is a weighting argument used to balance the data in the cases when different amplitudes are unequally represented in the target dataset.}
This function allows us to estimate the shear required for transforming each profile in the large dataset based on the saccadic amplitude (Figure~\ref{fig:lambda_fit}). Having the shearing factor for each amplitude $\alpha$, we apply shear $f(\alpha)$ to all individual profiles to form the new dataset. We compute such datasets for horizontal and vertical saccades, as well as for each user separately. 


\paragraph{Model shear}
It is possible to apply the shearing operation directly to an existing saccade prediction model as long as the individual saccade or mean saccade profiles can be recovered. Example of such a model is the one proposed by Arabadzhiyska et al.~\shortcite{Arabadzhiyska2017}. The model provides a mapping from time and displacement pair $(t,d)$ to the predicted saccade amplitude $\alpha$. Since the model is represented directly by the $(t_i,d_i,\alpha_i)$ triplets, the 
\begin{wrapfigure}{r}{0pt}
          \centering
          \includegraphics[width=0.65\linewidth]{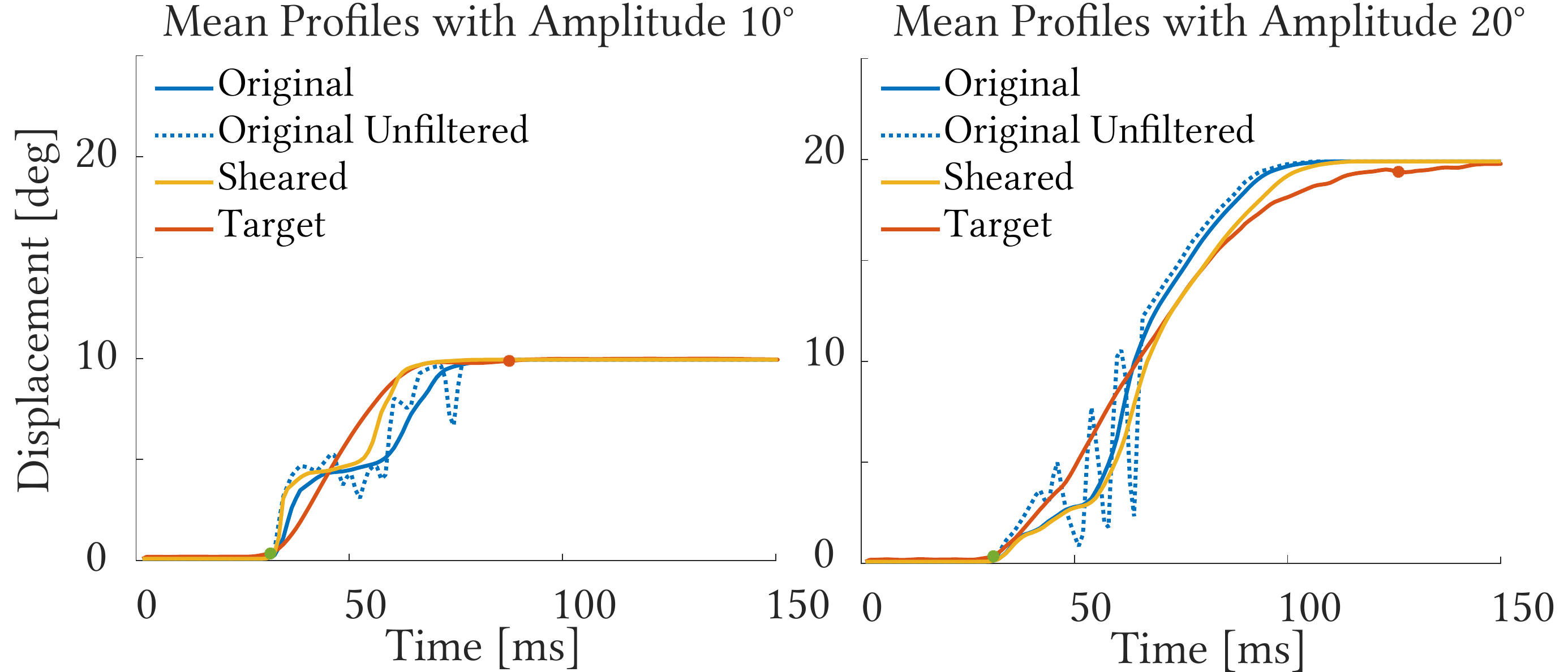}
          \captionof{figure}{Two examples of shearing saccade profiles, recovered from a model, to match the mean target profiles. \revcorr{The target on the left represents a category with faster saccades than the original recovered profile. The target on the right represents a category with slower saccades.}}
          \label{fig:model_shift}
    \end{wrapfigure}
mapping can be inverted by fixing $\alpha_i$ and treating the corresponding $(t_i,d_i)$ sequence as a displacement profile for a saccade with $\alpha_i$ degree amplitude. Because the model is represented by a discrete number of $(t_i,d_i,\alpha_i)$ sample points, we propose to use a linear interpolation on the displacement values to obtain saccade profiles sampled at regular, one-millisecond, intervals.
The blue dotted lines in \refFig{model_shift} show examples of the displacement profiles obtained using this procedure. Unfortunately, the profiles are often noisy, which prohibits a direct application of the shear with a satisfactory performance. \revcorr{ For this reason, as well as to prevent the occurrence of any aliasing, before shearing, we denoise the profiles by first applying a median filter with a window size of $15\,\ms$ followed by a Gaussian filter with a window size of $5\,\ms$ for smoothing (\refFig{model_shift}, blue solid lines). The values of the window size were chosen heuristically as the smallest values producing stable results.} After shearing the individual saccade profiles (\refFig{model_shift}, yellow lines), the new triplets $(\hat{t_i},d_i, \alpha_i)$ can be used to create a new model. Note that the shearing operation affects only timestamps, and the other components of the triplets do not change. In the particular case of the model of Arabadzhiyska et al.~\shortcite{Arabadzhiyska2017}, it is enough to resample the data to be uniformly sampled in time and displacement domain. We, therefore, apply linear interpolation to obtain $(t_i,d_i, \hat{\alpha_i})$ triplets, where $t_i$ and $d_i$ are sampled at the intervals of the original model, and $\hat{\alpha_i}$ is the new prediction of the saccade amplitude.
%
%

\subsection{Results}
\label{sec:results}

In our analysis, we consider both \textit{data shear} and \textit{model shear} strategies described in Section \ref{sec:application} to update saccade datasets and prediction models. We analyze the effectiveness of these strategies in two different experiments. In the first one, we show an application of shearing operation to update the existing dataset and prediction models for improved predictions when saccade orientation changes (horizontal vs. vertical). In the second experiment, we demonstrate the application of shearing operation to create user-specific models, aka personalization.

We compute our results on the saccade dataset and model of Arabadzhiyska et al.~\shortcite{Arabadzhiyska2017}, which includes 6600 saccade profiles collected from 22 participants (300 saccades for each participant). The amplitudes of saccades are evenly distributed in the range of $5 \degree - 45 \degree$. To customize the models for vertical and horizontal saccades, we classify saccades into horizontal and vertical categories depending on their orientation (with +/- 15 degrees allowance around the corresponding orientation). It is important to mention that, while the amplitude distribution across participants is balanced due to the experiment design, this is not the case for the orientation. Due to the aspect ratio of the screen (16:9), the amplitudes of vertical saccades are limited to the range of $5 \degree - 22 \degree$ while horizontal saccades have amplitudes up to $40\degree$. Moreover, horizontal saccades are more frequently represented in the dataset, constituting $30\%$ of the collected data, compared to $5\%$ for the vertical saccades.

\begin{table}
\vspace{0.5em}
\caption{\revcorr{Short descriptions of the four models that we compare in \refSec{results}. Each model is created following the procedure described by Arabadzhiyska et al.~\shortcite{Arabadzhiyska2017}, either using their full dataset or a subset of it that includes a single category of saccades (\refTbl{groups}). For \textit{data shear} we modify the dataset before creating the model and for \textit{model shear} we first create the model and then modify it to match a specific subset.}  }
\label{tbl:models}
\begin{center}
\begin{tabular}{|p{0.1\linewidth}|p{0.1\linewidth}|p{0.1\linewidth}|p{0.6\linewidth}|} 
 \hline
 Model & Original dataset & Target dataset & Model description \\ 
 \hline\hline
 Average & Full & - & The model is created using the original full dataset. \\ 
 \hline
 Model Shear & Full & Subset & The model is first created using the original full dataset, and then modified to match a specific subset of it.  \\
 \hline
 Data Shear & Full & Subset & The model is created from an augmented full dataset, modified to match a specific subset of the original dataset. \\
 \hline
 Customized & Subset & - & The model is created from a specific subset of the original dataset.  \\
 \hline
\end{tabular}
\end{center}
\end{table}

The baseline for all of our comparisons consists of two models. The first one is the \textit{average model} from Arabadzhiyska et al.~\shortcite{Arabadzhiyska2017}. It is derived from their dataset and is based on the interpolation of the collected data. We include this model in our comparisons because it provides a good balance between accuracy, performance, and data volume requirements. However, it accounts only for the variance in saccade profiles due to changes in the amplitude and it does not account for any additional factors that we considered in our paper (Section~\ref{sec:experiment_results}). The second model is the so-called \textit{customized model}, which is derived following the computation of the \textit{average model}, but using a subset of the data corresponding to a specific category of saccades (e.g., for horizontal or vertical saccades). \revcorr{\refTbl{models} gives a short summary of the four models that we compare in this section.}

In the first experiment, we computed the \textit{customized model} for the two categories of orientation (horizontal and vertical) separately. The number of saccades in each category was sufficient to properly train these models. Later, we used \textit{data shear} and \textit{model shear} as described in \refSec{application} to compute two alternative models and compare them with the \textit{customized models}. For \textit{data shear}, we sheared the displacement profiles of all saccades from the dataset, irrespectively of their orientation, according to the shear factor computed by using preselected horizontal and vertical saccades as target. For \textit{model shear}, the shearing factors were computed based on the comparison of the original model from Arabadzhiyska et al.~\shortcite{Arabadzhiyska2017} and the subsets of vertical and horizontal saccades. As for the second experiment, we followed a similar procedure to evaluate the performance of the shearing operation for personalizing the models, but in that case, saccades of a particular user were selected to compute the shearing factors.
    
Figure~\ref{fig:figures/direction-shift-error} presents the performance of different models tailored to the orientation of the saccade. The figure presents both the \revcorr{mean absolute} error (left), as well as the \revcorr{mean absolute} error for predictions made at a specific moment during the saccades (right).
\myfigure{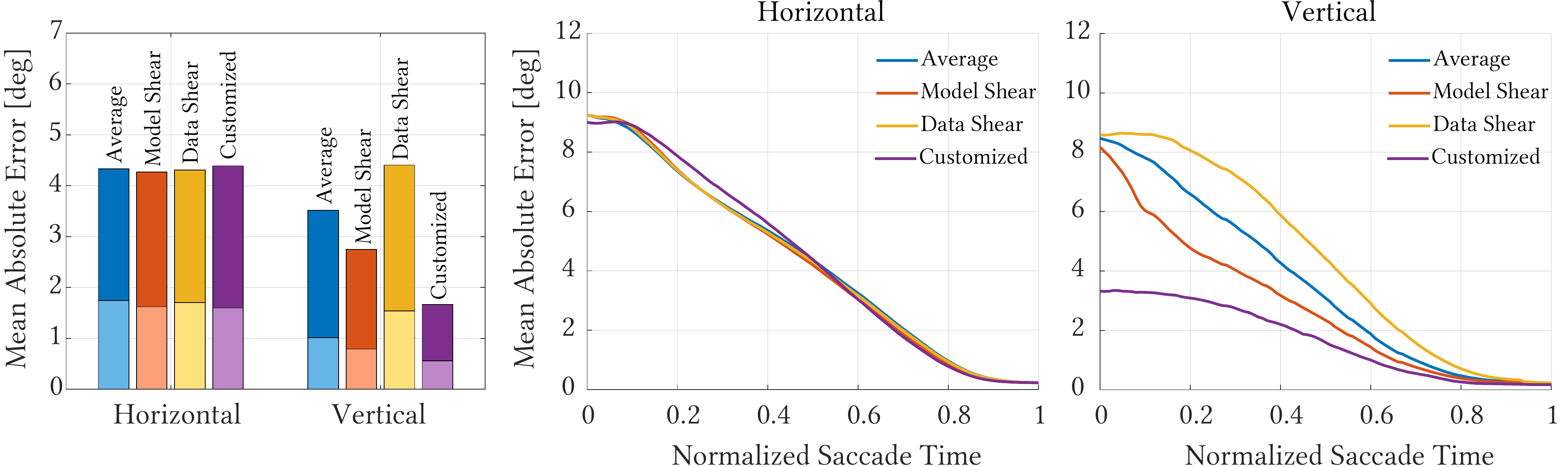}{
The figure presents the performance of differently derived models for horizontal and vertical saccades. The left-most plot presents aggregate mean errors for predictions made for the entire saccade duration (height of the bar) and for the second half of the duration (light segment). The two other plots present the error as a function of duration of the saccade, i.e., at which stage of the saccade the predictions was performed. While the \textit{customized model} performs best, the \textit{sheared model}, which requires significantly lower number of saccades for training, performs better than the \textit{average model}, which does not account for the orientation of the saccade.}
The performance of the horizontally oriented \revcorr{\textit{data shear} and \textit{model shear} models} is indistinguishable from the \textit{average} one. We attribute the lack of an effect to the predominance of the horizontal saccades in the dataset, and consequently, better prediction of these saccades. In comparison, the prediction for vertically oriented saccades greatly benefits from a vertically oriented models. It is important to mention here that the \textit{customized model} greatly benefits from the significantly lower range of amplitudes in the set of vertical saccades. More precisely, the range of the horizontal saccades is double the one of the vertical saccades due to the dimensions of the display used for the data collection \cite{Arabadzhiyska2017}. While reducing the training and testing range of saccades' amplitudes improves the prediction as the error is bound to this range, the model is limited to shorter saccades. In contrast, the models derived using \textit{model shear} and \textit{data shear} support the larger range of amplitudes represented in the original dataset.

\begin{wrapfigure}{r}{0pt}
      \centering
      \includegraphics[width=0.35\linewidth]{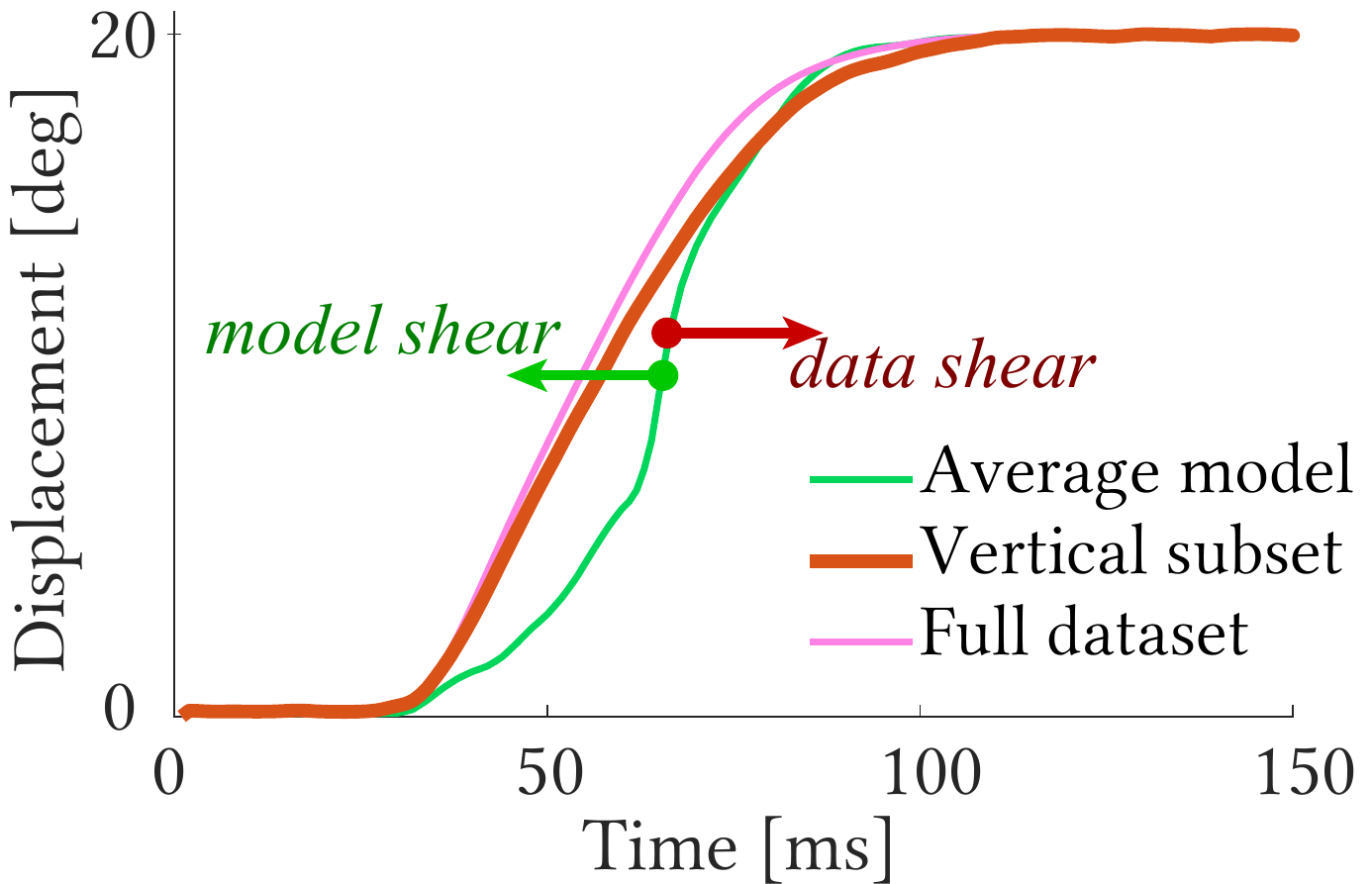}
      \captionof{figure}{
      \revcorr{Comparison of the mean saccade profiles from the full dataset (pink) and from the vertical subset (orange) with a recovered one from the \textit{average model} (green). The difference between the two mean profiles indicates slower performance of the vertical saccades with respect to the rest but the sample profile recovered from the average model does not reflect it.}
      }
      \label{fig:model_bias}
\end{wrapfigure}

While both \textit{model shear} and \textit{customized model} provided a better performance for the vertical saccades than the \textit{average model}, surprisingly, the \textit{data shear} did not improve the model. To understand the reason behind it, we analyzed the \revcorr{mean saccades profiles from the full dataset and from the vertical subset}, as well as the cross-section of the original average model \cite{Arabadzhiyska2017} in \refFig{model_bias}. \revcorr{When comparing the the mean saccade profiles representing the full dataset and the vertical subset, the first profile requires shearing to the right to match the second. This is expected as the vertical saccades are slower (Section~\ref{sec:experiment_discussion}). However, the cross-section of the model exhibits the opposite behavior, i.e. it requires shearing to the left to match the vertical saccades profiles.} When applying the \textit{model shear}, the shearing computed based on the vertical saccade profiles and the model results in the model shearing to the left (green arrow), hence, better aligning with the vertical saccades and reducing the error. 
\revcorr{However, shearing all the profiles in the dataset according to the difference between their representative mean profile and the vertical mean profile, i.e., \textit{data shear}, leads to a sub-optimal shear of the model to the right (red arrow), hence, increasing the prediction error, i.e., worse alignment with the vertical saccades profile.} This demonstrates that although \textit{data shear} can perform a correct transformation to the individual profiles, it cannot account for the built-in biases in the model. In this case, this leads to a lack of improvement when \textit{data shear} is followed by the model computation. Conversely, the \textit{model shear}, which computes the shearing factor based on the model, can account for biases in the model and improve the prediction.
%
%
%
\myfigure{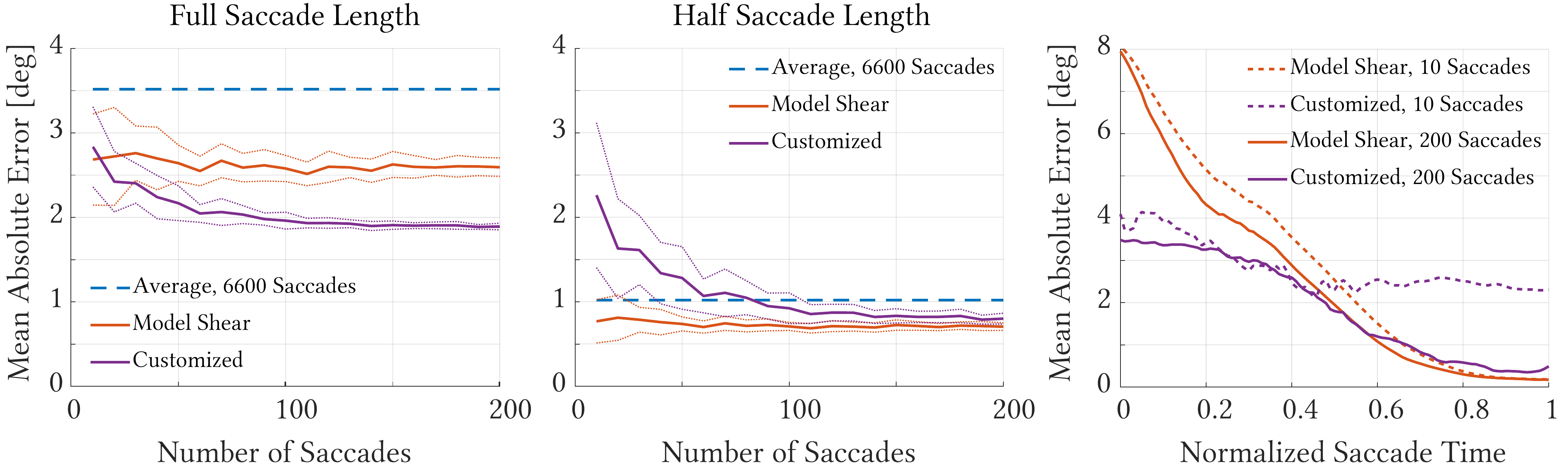}{Performance comparison for different models as a function of the number of saccades used for their computation. While the plot on the left shows the average error of the prediction for the full length of the saccade, the center plot shows the error for the prediction during the second half of the duration. The solid lines are the means computed using bootstrapping with $20$ repetitions, the dotted lines are the corresponding standard deviations. The plot on the right compares the average error of the prediction at any point during the saccade when using 10 and 200 saccades for training the models.}

The great potential of our shearing strategy lies in the fact that it may allow for training models using significantly lower number of samples than it is required for training \textit{customized models}. To verify this, we analyzed the performance of our shearing strategy for different numbers of saccades (Figure~\ref{fig:figures/personalized_vs_model}). To this end, we divided the dataset of vertical saccades into training and testing sets which consist of 200 and 150 saccades, respectively. By considering different number of saccades (x-axis in the plot) from the training set for computing the shearing factor and the new model, we analyzed the resulting mean absolute error of the prediction. We compared this \textit{model shear} strategy, to the straightforward computation of the model based on the smaller number of training saccades (\textit{customized model}). As expected, when the number of considered saccades is large, the improvement from our shearing technique may be limited. However, we can achieve a better prediction performance, in the presence of significantly lower number of saccades. This is particularly visible for the prediction in the second half of saccade duration, which is critical for techniques such as foveated rendering, where the sensitivity of the visual system is gradually restored towards the end of the saccade (Section~\ref{sec:saccadic_supression}). This can be in particularly observed in the right plot in Figure~\ref{fig:figures/personalized_vs_model}, where the error is analyzed for predictions made at different points of the saccades' duration. It can be observed that the \textit{customized model} trained on a low number of saccades retains the high error throughout the entire duration of the saccades. In contrast, the error for the model trained using our method drops significantly towards the end of the saccades.

In Figure~\ref{fig:figures/users-shift-error}, we provide the mean absolute error of predictions obtained from different models for personalization. We observe that for many participants (e.g., users 4, 12, 15, and 21) the prediction performance of models follow an expected pattern, where \textit{customized model} has the best performance due to the availability of full data used to calibrate such a model. \textit{Data shear} and \textit{model shear} provides the best prediction performances after \textit{customized model} and they are suitable for improving existing dataset or model prediction performances without large data collection requirements for personalization. The \textit{average model} performs least favorably due to the lack of user-based adjustments in saccade displacement profiles. Nevertheless, using a limited dataset for training prediction models is more prone the noise inherent to data. We observe that for some of the participants (e.g., users 7, 8, and 14) \textit{model shear} performs more favorably than \textit{data shear} and we attribute this observation to the model adjustments in \textit{model shear} that are more robust against noise. In some of the cases (e.g., users 3, 18, 19, and 20), \textit{data shear} and \textit{model shear} have a performance level close to that of the \textit{average model}. We believe that for those users, the personalization does not offer a high level of improvement in the performance. However, we observe that the general behavior of mean absolute errors favors the use of \textit{data shear} and \textit{model shear} for improving prediction performance without the cost of collecting a large amount of training data.
\myfigure{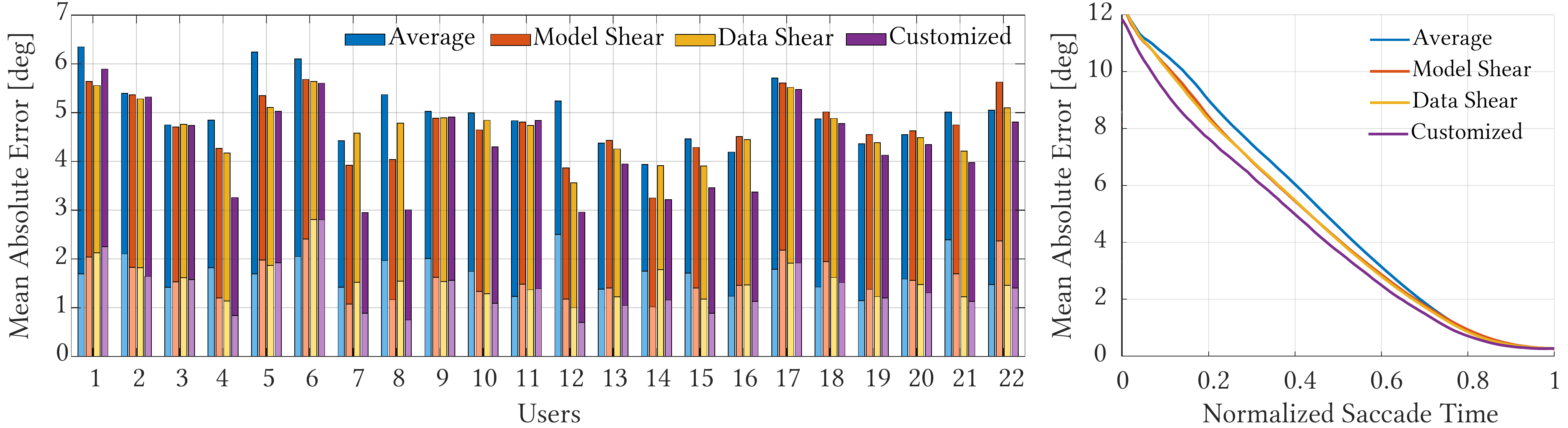}{The figure presents the performance of different models for each user. The bar plot on the left shows aggregate mean absolute errors for the saccade amplitude predictions. The height of the bars represents the mean error measured for whole duration of the saccade while the segments shaded with lighter colors represent the mean error measured in the second half of the saccade duration. The line plot on the right shows the mean absolute error as a function of point in time when the prediction was made during the saccade. The \textit{customized model} gives the best performance, followed by \textit{model shear} and then the \textit{average model} (please see the text for details). \textit{Model shear} mostly has a good prediction performance for the users, for whom the \textit{customized model} also performs well.}

Based on the above experiments, we conclude that both the \textit{data} and \textit{model shear} are viable solutions for extending and improving saccade prediction models to account for effects analyzed in Section~\ref{sec:experiment_results}. The important difference between them lies in how they can correct model biases. While the model shear is capable of correcting them, adjusting the data using \textit{data shear} is not. Therefore, the success of the \textit{data shear} is influenced by the quality of the prediction model built upon it.      

\section{Conclusion and Future work}

In many applications, such as foveated rendering, the latency poses significant challenges. Improving hardware solutions is one path for improving the performance of the techniques that benefit accurate gaze information. However, it has been demonstrated that latency problems can also be addressed by building efficient and accurate predictive models for fast eye movements \cite{Arabadzhiyska2017}. In this work, we go beyond existing models and analyze factors that should be accounted for when building such methods. We first demonstrate that factors, which were previously not considered explicitly, such as the orientation of the saccade, depth change, or initial smooth pursuit eye motion, affect the saccade profiles. Then, we propose a technique that allows extending previous models and datasets to train them to handle the additional effects while limiting the number of collected data in user experiments. We argue that this is critical for building comprehensive models for saccade prediction. The key to our technique is the proposed shearing operation which adapts previously derived models. This low parameter transformation acts as a regularization for smaller, possibly more noisy datasets. In this work, we demonstrated the performance of the method on training personalized models and models for horizontal and vertical saccades. In the future, the method can be used to train more comprehensive models addressing a continuous range of orientation, depth changes, user-specific factors, and possibly other factors using a lower number of input saccades. We also believe that the low number of parameters of the shear transformation will allow creating models that will adapt on the fly to the user without the additional need for calibration. Finally, our method can be seen as a data augmentation technique for machine learning techniques, such as \cite{morales2018}. While the current inference times do not meet the low latency demand of the state-of-the-art head-mounted displays, such techniques can provide acceptable performance and higher accuracy prediction in the future. In this context, our method can significantly limit the amount of data required for training such models facilitating the development and application of these techniques.

\bibliographystyle{ACM-Reference-Format}
\bibliography{main}

\end{document}